# Prototype development and validation of a beam-divergence control system for free-space laser communications

**Alberto Carrasco-Casado[1*], Koichi Shiratama[1], Dimitar Kolev[1], Phuc V. Trinh[1], Tetsuharu Fuse[1], Shingo Fuse[2], Koji Kawaguchi[2], Yusuke Hashimoto[2], Masamitsu Hyodo[2], Takashi Sakamoto[2], Terufusa Kunisada[2], Morio Toyoshima[1]**

[1]National Institute of Information and Communications Technology (NICT), Tokyo, Japan

[2]Tamron Co., Ltd., Saitama, Japan

\* Corresponding Author: alberto@nict.go.jp

## Abstract

Being able to dynamically control the transmitted-beam divergence can bring important advantages in free-space optical communications. Specifically, this technique can help to optimize the overall communications performance when the optimum laser-beam divergence is not fixed or known. This is the case in most realistic space laser communication systems, since the optimum beam divergence depends on multiple factors that can vary with time, such as the link distance, or cannot be accurately known, such as the actual pointing accuracy. A dynamic beam-divergence control allows to optimize the link performance for every platform, scenario, and condition. NICT is currently working towards the development of a series of versatile lasercom terminals that can fit a variety of conditions, for which the adaptive element of the transmitted beam divergence is a key element. This manuscript presents a prototype of a beam-divergence control system designed and developed by NICT and Tamron to evaluate this technique and to be later integrated within the lasercom terminals. The basic design of the prototype is introduced as well as the first validation tests that demonstrate its performance.

## 1   Introduction

The main advantage of free-space optical communications is the capacity to transmit very-narrow beams of light by using laser sources and collimation optics. This way, the communication signals can cover large distances with small losses because of the small achievable divergences. Dynamically controlling the beam-divergence angle of the transmitted signals has the potential to optimize the communications performance by increasing the optical-power density reaching the receiver. The reason is that there is an optimum value for the beam divergence depending on a variety of factors that cannot be accurately foreseen or that can change with time, making this optimum value vary. This is the case in most realistic space laser communication systems, since the optimum beam divergence depends on multiple factors that vary with time, such as the link distance, or cannot be accurately known, such as the actual pointing accuracy.

The traditional approach is to fix a beam divergence that matches the estimated pointing accuracy leaving a supplementary safety margin and choose some distance value that guarantees sufficient link availability. Since a trade-off is required, this strategy can lead to a very inefficient use of the system resources. If the system is designed to work in the worst case with confidence, i.e. wide beam divergence, then the communication performance will be significantly degraded or there will be an overallocation of resources. If the system is designed to optimize the onboard resources at the cost of increasing the risk of success, i.e. narrow beam divergence, then the link availability will be severely compromised or become inexistent. The capability to control the beam divergence in real time allows to avoid this suboptimum compromise and instead be able to adapt the system resources to a wide variety of varying conditions. This manuscript introduces the prototype of a beam-divergence control system which was developed to assess the aforementioned technique. The results of the first validation tests are shown as well, and they verify the required performance for this system in order to be integrated in real operational systems.

To illustrate the potential of the proposed system in a realistic scenario, as an example, it is possible to imagine a small lasercom system onboard a CubeSat relying on the attitude determination and control system (ADCS) for its pointing. Typically, the real CubeSat's pointing accuracy cannot be predicted precisely because it has a multifactor dependency, e.g., satellite's moment of inertia, presence and varying condition of deployable elements, total mass and its distribution, or even irregular ambient light impacting the star trackers. Taking a real case as an example, in the NASA-JPL's ASTERIA CubeSat mission, it was concluded that a pointing-accuracy improvement of almost 50 times could be achieved compared to its nominal specification of 0.021° (3σ in 3 axis) after in-orbit calibration and removal of bias and drifts [1]. The pointing loss $L_p = 10^{-2\beta^2}$, where $\beta = 2\sigma_p/\theta_d$, can be used to evaluate the relation between pointing accuracy $\sigma_p$ and beam divergence $\theta_d$. For a given $\sigma_p$, there is an optimum divergence $\theta_d \approx 5\sigma_p$ [2]. If a conventional lasercom system had to be designed for the previous CubeSat using the nominal ADCS even with no margin and using the optimum $\sigma_p/\theta_d$ ratio, the beam divergence should be set at $\theta_d \approx 1.8$ mrad. However, if beam divergence could be reduced, such satellite could support a divergence as small as 39 μrad. Considering that the transmitting gain can be calculated as $G_T = 16/\theta_d$ [3], beam-divergence optimization could provide an additional gain of 17 dB in this example achievable by a 4-cm aperture, or a gain of 13 dB for the specific 2-cm aperture system proposed in this manuscript.

## 2 Design of the beam-divergence control prototype

NICT is currently developing versatile lasercom terminals with the goal to fit a wide range of circumstances without the need of ad-hoc design or customization [4]. This approach was considered to be strategic in the plan to achieve functional terminals that can be truly operational in real communication networks. Beam-divergence control is a key part of these terminals to optimize the use of the available resources to the varying conditions. Such a system can allow to optimize the link performance in every platform and scenario by achieving the narrowest divergence angle that the system resources can support in any given condition.

The requirements of the first prototype were defined as a standalone system that could be later integrated in a small lasercom terminal operating from LEO to ground and function as an adaptive-transmitter subassembly. Since the system was conceived to be compatible with CubeSats as the strategy to enable its integration with any other kind of platform, the maximum optical aperture was assumed to be 2 cm. Such aperture size can be easily accommodated by a transmissive optical system made up with lenses, avoiding the use of reflective assemblies in which the secondary mirror blocks the central area where the highest intensity is concentrated. In transmissive systems with no central obscuration, the optimum truncation of the gaussian beam is a ratio between the aperture and the beam size of 1.12 [5] to yield the maximum on-axis gain. Therefore, the $1/e^2$ gaussian beam size was designed to be 1.78 cm.

At a wavelength of 1550 nm, a 1.78-cm gaussian beam transmitted through a 2-cm aperture produces a beam divergence of 90 µrad (FWHM). This is the minimum divergence that can be set with the above conditions, hence the assumption for the collimated-beam's divergence. This divergence allows to close a downlink from the LEO orbit with a 5-dB link margin based on IM/DD up to 10 Gbit/s at the closest approach (600 km) or 2.5 Gbit/s at the longest approach (1200 km). By being able to increase the beam divergence, it is possible to design the link for such an ideal case (the footprint becomes only 54 meters when reaching the receiver on the ground at the closest approach) because by increasing the beam divergence, the system can adapt to worse conditions. For example, a stronger-than-expected platform's vibration or an unforeseen degradation of the satellite's attitude control. For this link budget, a 2-W transmitted optical power was assumed, by using a miniaturized EDFA (compatible with CubeSat's platforms) that was already developed and qualified for space by NICT [6], although there is nothing in the current design that prevents from using a higher power. For the ground segment, a small 35-cm telescope was assumed by using a transportable optical ground station that was already developed by NICT as well [7].

The maximum divergence was designed to be 5 mrad, approximately 50 times the minimum divergence. Such a wide divergence makes it possible to support CubeSat's ADCS with pointing accuracies in the 0.1° range, which cost can be several times cheaper than high-end ADCS such as the one mentioned in the introduction. Although such a wide divergence will not be normally necessary for communications-performance optimization during regular operations, there are other scenarios where such a wide divergence is advantageous. For example, to cover a wider range including shorter distances. This can happen in intersatellite links between nodes within a LEO constellation, where the distance can span from tens to thousands of km. Another example is for rapid beam acquisition by using the same lasercom transmitter for beacon and communications. Additionally, such a feature can guarantee agile beam acquisitions in LEO constellations where terminals move fast with respect to each other, with short contact time, which gives little time for beam scanning to find the counter-terminal. Furthermore, as explained in section 3, such divergence range can be achieved with a very-small linear motion of the moving-lens group. And lastly, a wide range of divergence angles makes it possible to use the same device in many different scenarios without customization.

There are two different ways to increase the divergence of a laser beam for given values of output aperture and beam truncation, i.e., to reduce the size of the collimated beam, or to modify the beam-focusing characteristics. For lasercom applications, both methods produce equivalent results in the far field, provided that they produce laser beams with an equivalent wavefront quality. The second method was used in the presented system because it allows a simpler implementation. For example, the maximum-to-minimum beam-size ratio is 6 times for the presented method, but it would be of 55 times for the first method, meaning that the maximum divergence would require to collimate a 325-µm beam, which is much more challenging, since it requires a more-precise control of the position of the moving-lenses group, and it is more susceptible to suffer from optical aberrations.

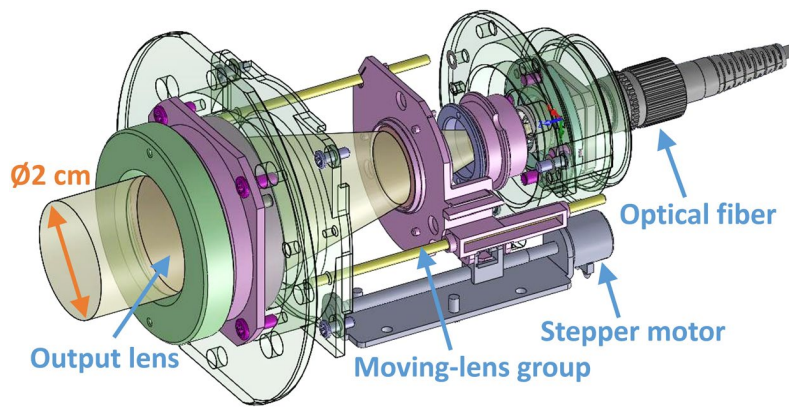

Figure 1. Basic design of the beam-divergence control system.

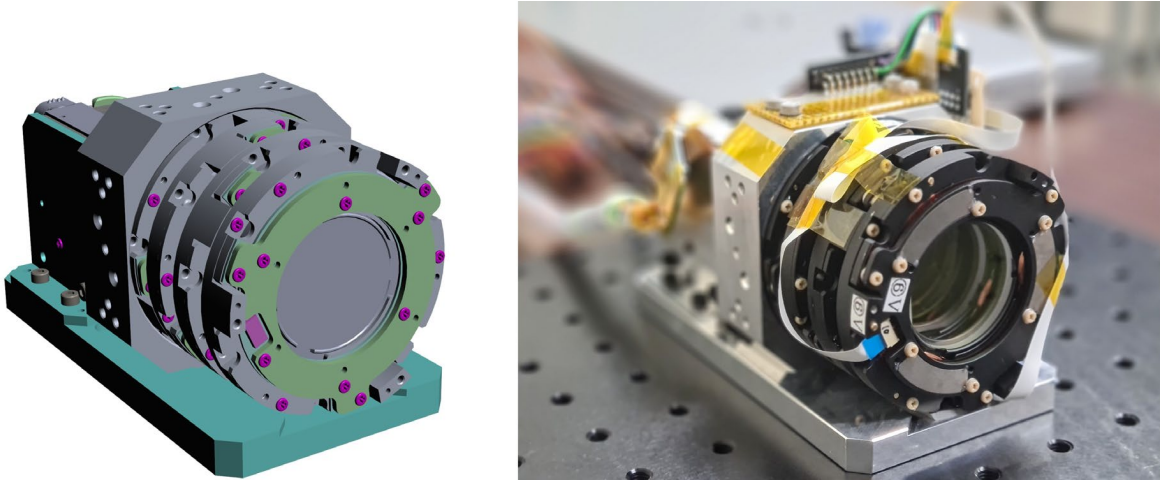

Figure 2. 3D model of the beam-divergence control system (left),
and first prototype of the beam-divergence control system (right)

The basic design of the beam-divergence control system is shown in fig. 1. The moving-lens group is governed by a stepper motor connected through a rail system leveraged from the Tamron's expertise in variable-zoom camera lenses. The optical fiber carrying the communication laser is connected through an FC/APC connector in the system's input, and the output lens has a clear aperture of 2 cm. An experimental anti-vibration and beam-steering mechanism was also designed and developed in order to remove platform vibration up to 100 Hz (as most of the CubeSat's vibration stays below tens of Hz [8]) and also in order to correct misalignments and perform fine pointing with a total range of ±100 µrad in the 2 axis. This system is based on pass-through on-axis rotational Wedge prisms with the goal to avoid coma aberration when the beam is steered. Fig. 2 shows the full prototype including the beam divergence control and the anti-vibration and beam-steering system. Although the first prototype has an external dimension of 85×62×58 mm, since it was designed for functional-verification purposes, a 2× size reduction is expected for the final miniaturized version.

## 3    Verification of the beam-divergence control prototype

This section describes the test campaign that was carried out to validate the prototype's performance (see table 1 for a summary). The first test was to confirm the designed insertion loss, which was initially required to be less than 0.5 dB (transmittance higher than 89.5%). When using a 1550-nm laser source, the transmittance of the beam-divergence control alone accounts for 99.392% and 99.270% for the full system including the anti-vibration subsystem, thus 0.026 dB and 0.032 dB loss respectively, keeping similar values for the full C-band. Regarding the optical power that the device allows, the initial requirement was 2 W, as explained in section 2, and after a long exposure to this intensity level, no degradation was observed when analyzing the wavefront quality before and after the experiment. The maximum power consumption of the optomechanical module is 0.7 W, although in its normal operation this power would only be required for the short time it takes to set the desired divergence, which is in the order of milliseconds.

Table 1. Beam divergence control characteristics.

| Item | Value |
|---|---|
| Optical aperture | 20 mm |
| Beam waist ($1/e^2$) | 17.8 mm |
| Spectral band | C band |
| Divergence range (FWHM) | 90 - 6250 µrad |
| Divergence setting speed | $\geq 10$ µrad/ms |
| Divergence setting accuracy | $\leq \pm 1\%$ |
| Optical axis stability | $\leq \pm 5\%$ |
| Vibration isolation | $\leq 100$ Hz |
| Optical power | $\leq 2$ W (CW) |
| Total insertion loss | $\leq 0.1$ dB |
| Qualified environment | Thermal vacuum |
| Temperature range | -30 to 60°C |

The most important validation test refers to the beam-divergence control itself, how well the prototype can set the desired divergence angle, what wavefront quality can be achieved throughout the entire range, and how repeatable the beam divergence is. A Shearing interferometer was used for the first two verifications since this technique allows to measure the wavefront's curvature of monochromatic sources and to evaluate the optical aberrations [9]. A wedged lateral Shearing interferometer was used with the prototype to find the collimated-beam position as well as to verify the absence of aberrations in the transmitted beam. Some captions of the IR camera used after the interferometer are included in Fig. 3, where the absence of aberrations throughout the full operation range can be appreciated in the parallel fringes.

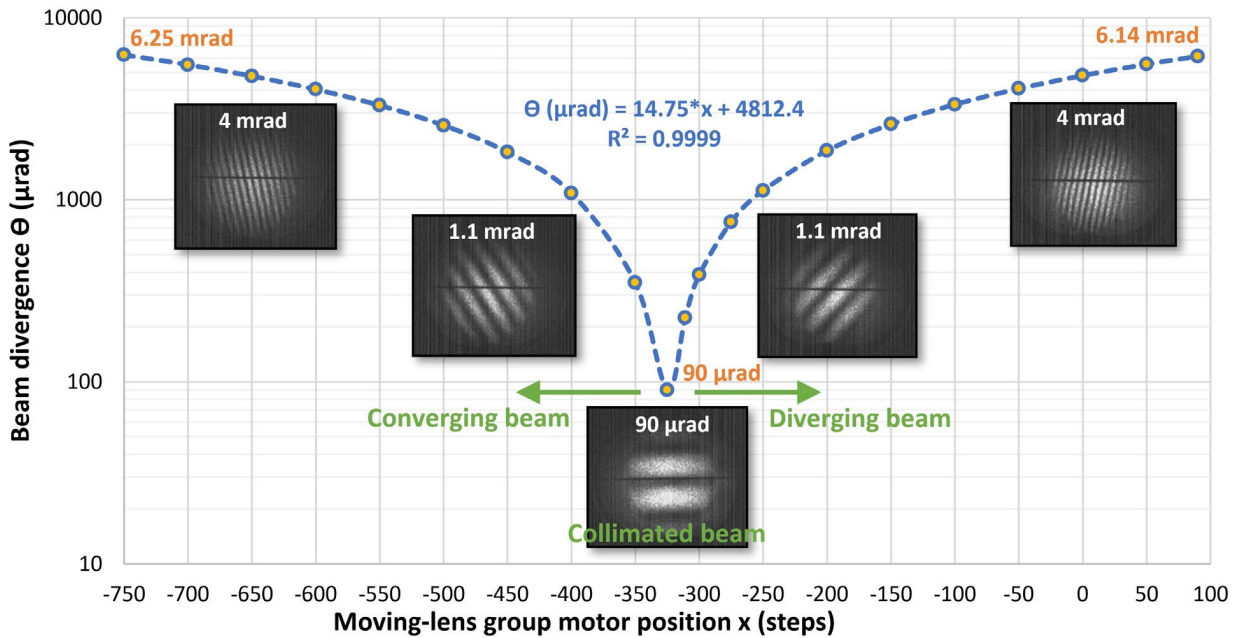

Figure 3. Beam divergence of the prototype depending on the position of the moving lens.

The actual beam divergence that the prototype was capable to set over a 1550-nm laser was verified by using the following method. A 15-meter horizontal lane was built to measure the beam size in 4 different points: at 3 m (after the beam waist), at 5 m, at 10 m, and at 15 m. In each position, an infrared beam profiler designed for large beam size and high optical power was used to measure the $1/e^2$ spot diameter. The beam profiler LaseView-LHB-100-NIR-GigE allows to measure beam sizes up to 10×10 cm with a resolution of 800 µm, which is better than 4.5% of the smallest beam size. After averaging the measurements in each of the 4 positions, a linear relationship could be established between the lens position and the divergence angle with an $R^2$ coefficient of 0.9999 when fitting the linear regression with the experimental data (see fig. 3). The prototype allows setting divergences wider than the minimum one by making the beam diverge or converge, and although the maximum divergence was 5 mrad by design, the developed prototype allows setting up to 6.14 mrad in the diverging range and 6.25 mrad in the converging range. The moving-lens group achieves all this range within 3.5 mm of linear motion in each direction, taking 0.9 seconds in total, thus achieving an angular speed of 13.6 µrad/ms to go from any given divergence to a different value.

The minimum divergence could also be confirmed to be as designed, i.e. 90 µrad (FWHM), which is the absolute minimum angle that a 17.8-mm beam can diverge. However, for this divergence to be achieved accurately, it is necessary to measure the actual numerical aperture of the single-mode fiber and design the optics for this real value, otherwise there will be a degradation of the minimum achievable divergence due to the mismatch. For example, in the first iteration of the prototype's development, a difference of 2.62° between the real and the nominal fiber's numerical aperture led to a difference in the collimated beam's size of 3.5 mm (smaller than designed), which produced a 23-µrad wider divergence. In addition, due to the narrow angle when the beam is collimated and the limited angle resolution of the beam profiler, this specific measurement must be averaged over a number of samples to achieve a confident value. After averaging the measured beam divergence of the collimated beam, a deviation of 1.04% was found over the nominal 90-µrad value. Since this system was conceived to work with monochromatic laser sources, achromatic behavior was not required by design, presenting a dependence of beam divergence with wavelength, although with no wavefront degradation. When considering the full C-band, this dependence translates into a 10-µrad increase at 1530 nm and a 3-µrad increase at 1565 nm with respect to the optimized 90-µrad collimated beam at 1550 nm. For the maximum divergence of 5,000 mrad, it increases by 171 µrad for 1530 nm and 130 µrad for 1565 nm.

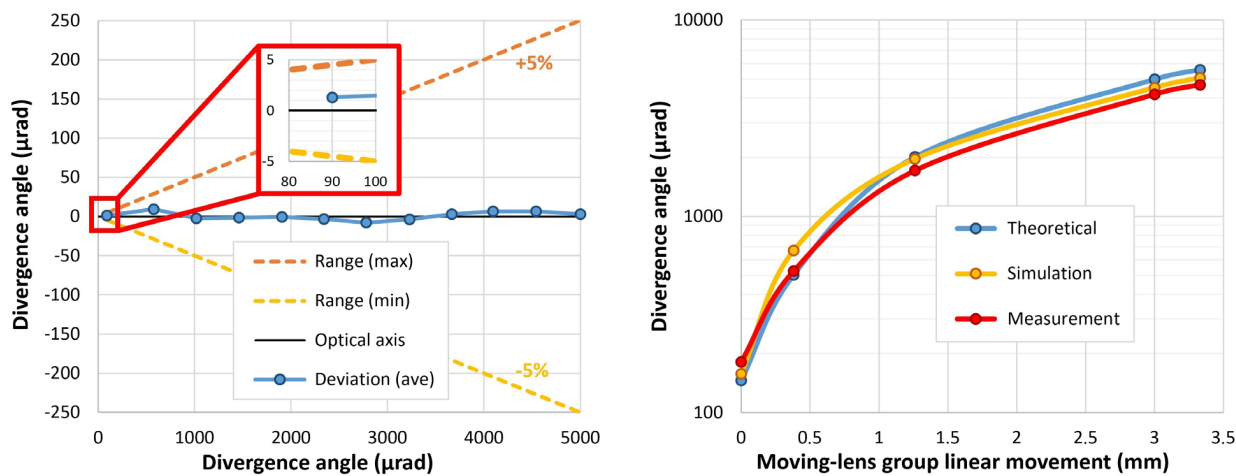

Figure 4. Prototype's optical-axis stability throughout the full divergence range(left), and comparison between theoretical design, simulation, and measurement (right).

A key feature of beam-divergence control is the optical-axis stability, especially when the beam is collimated. By design, the stability was required to stay within ±5% through the full range. For a 90-μrad collimated beam, this means that the optical axis cannot change by more than 4.5 μrad. To verify this feature, the deviation angle from the optical axis of the beam was measured at a distance of 15 meters from the prototype for 7 times for each of 12 different divergence values. As shown in the left side of fig. 4, the optical-axis stability was confirmed to be well within the required ±5% of each divergence value for all the measurements, being the collimated-beam case the most challenging one, but keeping an average axis deviation of 1.3 μrad at that divergence. The right side of Fig. 4 shows a comparison between the $1/e^2$ values of the theoretically-designed divergence depending on the linear movement of the moving-lens group, the simulated-designed divergence by using the Code-V optical design software, and the measured divergence at the first iteration before optimization. The figure shows a good fit between predicted and measured data, validating the approach of theory and simulation.

Since the beam-divergence control system was conceived for space-environment operation, preliminary space-qualification tests were carried out. By design, the prototype was designed to be compatible with vacuum and radiation environment, although the latter has not been tested yet. The prototype was tested in vacuum environment and with a thermal cycling ranging from -30°C to +60°C in 5 steps: -30°C, -20°C, 20°C, 40°C and 60°C, each for the minimum and maximum divergence. No change in the divergence angle or wavefront profile was observed under vacuum operation. However, significant differences were observed in the divergence angle depending on the temperature, although not in the wavefront profile. The biggest angle deviation happens with the collimated beam, where the divergence becomes 7.5 times wider than expected at -30°C and 4.7 times wider at 60°C (and only 1.2 times narrower at -30°C and 1.1 times wider at 60°C when setting a divergence of 5 mrad). Nevertheless, this dependence on temperature is linear, repeatable, and correctable well within the system's tuning range. Therefore, a lookup table can be created to apply the correction, although other strategies are being investigated as well to find an optimum solution.

## 4    Conclusion and future work

A prototype of a beam-divergence control system was designed and developed by NICT together with Tamron with the goal to introduce operation flexibility in the series of lasercom terminals that NICT is currently working on. This paper introduced the design concept, the first functional prototype, as well as the validation tests. The positive results obtained with the prototype have led to continue its development, and at the moment further efforts are being carried out to miniaturize the final device and to improve the overall performance when integrated in the real lasercom terminals.

## 5    Conflict of Interest

Authors Shingo Fuse, Koji Kawaguchi, Yusuke Hashimoto, Masamitsu Hyodo, Takashi Sakamoto and Terufusa Kunisada are employed by Tamron Co., Ltd. The remaining authors declare that the research was conducted in the absence of any commercial or financial relationships that could be construed as a potential conflict of interest.

## 6    References

[1] C. M. Pong, "On-Orbit Performance & Operation of the Attitude & Pointing Control Subsystems on ASTERIA," 32nd Annual AIAA/USU Conference on Small Satellites, SSC18-PI-34 (2018).

[2] K. Kiasaleh, "On the probability density function of signal intensity in free-space optical communications systems impaired by pointing jitter and turbulence," Opt. Eng., Vol. 33, No. 11 (1994).

[3] A. Carrasco-Casado, R. Mata-Calvo, Space Optical Links for Communication Networks, in: B. Mukherjee, I. Tomkos, M. Tornatore, P. Winzer, Y. Zhao (Eds.), Springer Handbook of Optical Networks, Springer Handbooks, Springer, Cham, 2020, pp. 1057–1103.

[4] A. Carrasco-Casado, "Free-space Laser Communications for Small Moving Platforms," M4I.1 (invited paper), Optical Fiber Communication Conference, San Diego, California, United States (2022).

[5] B. J. Klein and J. J. Degnan, "Optical Antenna Gain. 1: Transmitting Antennas," Applied Optics, Vol. 13, Issue 9 (1974).

[6] A. Carrasco-Casado, K. Shiratama, P. V. Trinh, D. Kolev, Y. Munemasa, Y. Saito, H. Tsuji and M. Toyoshima, "Development of a miniaturized laser-communication terminal for small satellites," IAC-21/B2/2, 72nd International Astronautical Congress, Dubai, United Arab Emirates (2021).

[7] Y. Saito, H. Takenaka, K. Shiratama, Y. Munemasa, A. Carrasco-Casado, P. V. Trinh, K. Suzuki, T. Fuse, Y. Takahashi, T. Kubo-oka and M. Toyoshima, "Research and Development of a Transportable Optical Ground Station in NICT: The Results of the First Performance Test," Frontiers in Physics – Optics and Photonics, Vol. 9 (2021).

[8] R. Antonello, F. Branz, F. Sansone, A. Cenedese and A. Francesconi, "High Precision Dual-Stage Pointing Mechanism for Miniature Satellite Laser Communication Terminals," IEEE Transactions on Industrial Electronics, Vol. 68, Issue 1 (2021).

[9] M. E. Riley and M. A. Gusinow, "Laser beam divergence utilizing a lateral Shearing interferometer," Applied Optics, Vol. 16, No. 10 (1977).